# Flow problem in three-dimensional geometry

Maxim Zaytsev[1)], Vyacheslav Akkerman[2)]

[1)] Nuclear Safety Institute of Russian Academy of Sciences

115191 Moscow, ul. Bolshaya Tulskaya, 52, Russia

[2)] West Virginia University Morgantown, WV 26506-6106, USA

**Abstract**

It is shown how a complete set of hydrodynamic equations describing an unsteady three-dimensional viscous flow nearby a solid body, can be reduced to a closed system of surface equations using the method of dimension reduction of over-determined systems of differential equations. These systems of equations allow determining the surface distribution of the resulting stresses on the surface of this body as well as all other quantities characterizing the hydrodynamic flow around it.

**Keywords**: flow problem, hydrodynamic flow, viscosity, solids, stress on the body surface, differential equation on the surface

## Introduction

In application, nonlinear equations of hydrodynamics are often one of the obstacles involved in the study of many phenomenas, such as turbulence, hydrodynamic discontinuities motion, movement of the fronts of chemical reactions, etc. [1-4]. (During their modeling in practical problems often one has to produce a very fine numerical grid in space and time, which requires large computing power and time-consuming) When modeling such equations in practical problem, one has to often produce a very fine numerical grid in space and time which is time consumming and requires large computing power hence rendering the methode as not practical [5]. In this regard, various ways of reducing the complete system of hydrodynamic equations in terms of a system of surface equations are of a large scientific interest. [6-11]. Such a procedure can reduce the dimension of the problem by one ($3D \to 2D, 2D \to 1D$), which significantly reduces the required computational power. In particular, the corresponding computer program could directly calculate the hydrodynamic discontinuities with viscosity, sound formation and other changes in the density of gases and liquids (using the information only on the surface). For example, the problem of describing the potential flow in a plane can be reduced to an integral equation on the boundary for a vanishingly small viscosity (Dirichlet, Neumann) [1, 4]. This equation relates the tangential and normal components of the velocity. If one of the components is known on the boundary of the body, one can define all the external flow.

In this paper we reduce the equations of a viscous medium "in the volume" to surface equations and demonstrate how the proposed method can reduce the dimension of the problems in viscous flow, which makes the equations easier to simplify the calculations in a variety of applications. General methods proposed for overdetemination of the Navier-Stokes equations in three-dimensional and two-dimensional case, which theoretically allows reducing them to a closed system of equations on any surface [12, 13]. No significant simplifying assumptions, which would limit the generality of the consideration of these phenomena, are considered.

1. **Formulation of problem**

Movement of solids in viscous media is characterized by the conditions on the boundary of a solid, where the particles stick to it. In other words, the rate of viscous medium on the surface of the body **u** is determined by the kinematic characteristics of the surface, for example, it can be expressed in terms of the speed of the body surface $V$. Force **F** acting on a unit surface area is [1]

$$F_i = P n_i - \sigma'_{ik} n_k \tag{1}$$

where **n** is normal to the surface, and

$$\sigma'_{ik} = \eta \left( \frac{\partial u_i}{\partial x_k} + \frac{\partial u_k}{\partial x_i} \right) \tag{2}$$

the viscous stress tensor and $\eta$ - coefficient of dynamic viscosity. Therefore, according to equation (1), in this case, for any point on the body surface in a Cartesian coordinate system $(\mathbf{n}, \boldsymbol{\tau}_1, \boldsymbol{\tau}_2)$ one has

$$F_n = P + 2\eta \left( \frac{\partial u_1}{\partial \tau_1} + \frac{\partial u_2}{\partial \tau_2} \right), \tag{3}$$

$$F_1 = \eta \left( \frac{\partial u_1}{\partial n} + \frac{\partial u_n}{\partial \tau_1} \right) = \eta \left( \omega_2 + 2 \frac{\partial u_n}{\partial \tau_1} \right), \tag{4}$$

$$F_2 = \eta \left( \frac{\partial u_2}{\partial n} + \frac{\partial u_n}{\partial \tau_2} \right) = \eta \left( -\omega_1 + 2 \frac{\partial u_n}{\partial \tau_2} \right) \tag{5}$$

(see Fig. 1). In addition, the force **F** relates to the velocity of the body $V$ via the laws of rigid body motion dynamics. Unsteady flow around fixed solids in viscous medium is different from the above by cancelling velocity conditions at the boundary [1]:

$$V = 0 \text{ и } \mathbf{u} = 0. \tag{6}$$

## 2. Overdetermination of Navier-Stokes equations

We confine ourselves here to an incompressible viscous flow in three-dimensional space in the gravity field **g**. To illustrate this, the Navier-Stokes equations take the form [1]

$$\frac{\partial \mathbf{u}}{\partial t} - \mathbf{u} \times \boldsymbol{\omega} + \nabla \left( P + \frac{1}{2} \mathbf{u}^2 \right) = -\nu \nabla \times \boldsymbol{\omega} + \mathbf{g}, \tag{7}$$

$$\boldsymbol{\omega} = \nabla \times \mathbf{u}, \tag{8}$$

$$\text{div}\,\mathbf{u} = 0, \tag{9}$$

where $\nu$ - is the kinematic viscosity coefficient. These equations allow us to find the unknown velocity vector $\mathbf{u}(\mathbf{r},t)$ and the pressure profile $P(\mathbf{r},t)$ at any given point and time in space, i.e. for any $(\mathbf{r},t)$. We introduce a "correction" $\boldsymbol{\alpha}$ to the vector **u** and "pseudo-density" $\rho(\mathbf{r},t)$ in order to generalize the system of equations (7) - (9) as follows

$$\frac{\partial (\mathbf{u}+\boldsymbol{\alpha})}{\partial t}-\left[(\mathbf{u}+\boldsymbol{\alpha})\times(\boldsymbol{\omega}+\nabla\times\boldsymbol{\alpha})\right]+\nabla\left(P+\frac{1}{2}\mathbf{u}^2\right)=\mathbf{g}, \qquad (10)$$

$$\frac{\partial \rho}{\partial t}+(\mathbf{u}+\boldsymbol{\alpha})\cdot\nabla\rho+\rho\cdot\mathrm{div}(\mathbf{u}+\boldsymbol{\alpha})=0, \qquad (11)$$

$$\frac{\partial \boldsymbol{\alpha}}{\partial t}=\left[\boldsymbol{\alpha}\times\boldsymbol{\omega}+\mathbf{u}\times(\nabla\times\boldsymbol{\alpha})+\boldsymbol{\alpha}\times(\nabla\times\boldsymbol{\alpha})\right]+\nu\nabla\times\boldsymbol{\omega}, \qquad (12)$$

$$\boldsymbol{\omega}=\nabla\times\mathbf{u}, \qquad (13)$$

$$\mathrm{div}\,\mathbf{u}=0. \qquad (14)$$

Actually, the equations (11) and (12) respectively determine $\rho(\mathbf{r},t)$ and $\boldsymbol{\alpha}$ but the equation (10) is resulting from Equations (7) and (12). Vector field $\mathbf{u}+\boldsymbol{\alpha}$ can be formally considered as *inviscid* hydrodynamic flow with density $\rho(\mathbf{r},t)$ and pressure $\left[P+\frac{1}{2}\mathbf{u}^2-\frac{1}{2}(\mathbf{u}+\boldsymbol{\alpha})^2\right]$.

We go one to one to the pseudo Lagrangian variables of initial position of the gas particles and time $(\mathbf{r}_0,t)$ (i.e. marking)

$$\frac{d\mathbf{r}}{dt}=\mathbf{u}(\mathbf{r},t)+\boldsymbol{\alpha}(\mathbf{r},t)+\lambda(\boldsymbol{\omega}(\mathbf{r},t)+\nabla\times\boldsymbol{\alpha}(\mathbf{r},t)),$$

$$\mathbf{r}=\mathbf{r}(\mathbf{r}_0,t)\ \text{и}\ \mathbf{r}_0=\mathbf{r}_0(\mathbf{r},t),\ \mathbf{r}_0\big|_{t=0}=\mathbf{r}. \qquad (15)$$

We chose $\lambda(\mathbf{r},t)$ so that

$$\lambda(\boldsymbol{\omega}+\nabla\times\boldsymbol{\alpha})\cdot\nabla\rho+\rho(\boldsymbol{\omega}+\nabla\times\boldsymbol{\alpha})\cdot\nabla\lambda=0. \qquad (16)$$

Hence, it can be shown that

$$\frac{\partial \mathbf{r}_0}{\partial t}+(\mathbf{u}+\boldsymbol{\alpha}+\lambda(\boldsymbol{\omega}+\nabla\times\boldsymbol{\alpha}))\cdot\nabla\mathbf{r}_0=0. \qquad (17)$$

Then, according to appendix A, from the equations (10) and (11) one yields the following expressions:

$$\frac{(\boldsymbol{\omega}+\nabla\times\boldsymbol{\alpha})\cdot\nabla\mathbf{r}_0}{\rho}=\frac{(\boldsymbol{\omega}_0(\mathbf{r}_0)+\nabla\times\boldsymbol{\alpha}_0(\mathbf{r}_0))}{\rho_0(\mathbf{r}_0)}, \qquad (18)$$

$$\frac{1}{\rho}\frac{\partial(x_0,y_0,z_0)}{\partial(x,y,z)}=\frac{1}{\rho_0(\mathbf{r}_0)}, \qquad (19)$$

where $\boldsymbol{\alpha}_0$, $\rho_0$, $\boldsymbol{\omega}_0=\nabla\times\mathbf{u}_0$ - initial distribution of the variables $\boldsymbol{\alpha}$, $\rho$, $\boldsymbol{\omega}$. Let $\boldsymbol{\alpha}_0\equiv 0$ and $\rho_0\equiv 1$. Then (18) and (19) can be transformed to the form [14]:

$$\omega_x+\nabla\times\alpha_x-\left|\frac{\partial \mathbf{r}_0}{\partial y}\ \frac{\partial \mathbf{r}_0}{\partial z}\ \boldsymbol{\omega}_0(\mathbf{r}_0)\right|=0, \qquad (20)$$

$$\omega_y + \nabla \times \alpha_y - \left| \begin{array}{ccc} \dfrac{\partial \mathbf{r_0}}{\partial z} & \dfrac{\partial \mathbf{r_0}}{\partial x} & \boldsymbol{\omega}_0(\mathbf{r_0}) \end{array} \right| = 0, \qquad (21)$$

$$\omega_z + \nabla \times \alpha_z - \left| \begin{array}{ccc} \dfrac{\partial \mathbf{r_0}}{\partial x} & \dfrac{\partial \mathbf{r_0}}{\partial y} & \boldsymbol{\omega}_0(\mathbf{r_0}) \end{array} \right| = 0. \qquad (22)$$

Actually, from three equations (20) - (22) only two are independent [14]. Formula (19) takes the form

$$\frac{1}{\rho} \frac{\partial(x_0, y_0, z_0)}{\partial(x, y, z)} = 1. \qquad (23)$$

Thus, we have obtained an overdetermined system of 16 differential equations (7) - (9) (11) (12) (16) (20) (21) (23) plus any of the equations from (17) and of 15 unknowns $\boldsymbol{\alpha}$, $\mathbf{u}$, $\boldsymbol{\omega}$, $P$, $\rho$, $\lambda$ и $\mathbf{r}_0$. Causation analysis shows that the system is consistent and valid [12-14]. It is possible to add to the equation (12) a nontrivial external force.

## 3. Viscous flow in a three-dimensional case

We derive a system of equations describing the unsteady flow around the fixed rigid body in three-dimensional incompressible viscous stream. Let us denote

$$S_1 = \alpha_x,\ S_2 = \alpha_y,\ S_3 = \alpha_z,\ S_4 = u_x,\ S_5 = u_y,\ S_6 = u_z,\ S_7 = \omega_x,\ S_8 = \omega_y,\ S_9 = \omega_z,$$

$$S_{10} = P,\ S_{11} = \rho,\ S_{12} = \lambda,\ S_{13} = x_0,\ S_{14} = y_0,\ S_{15} = z_0. \qquad (24)$$

Then, a system of 16 equations (7) - (9), (11), (12), (16), (20), (21), (23) plus any of the equations (17) can be written in form

$$H_k\left(S_v, \frac{\partial S_v}{\partial \mathbf{r}}, \frac{\partial S_v}{\partial t} \ldots \right) = 0,\ v = 1\ldots 15,\ k = 1\ldots 16. \qquad (25)$$

We now turn to the coordinate system $(\boldsymbol{\tau}_1, \boldsymbol{\tau}_2, \mathbf{n})$ at point $M$ (see Fig. 1). Then equations (25) can be written as

$$H_k\left(S_v, \frac{\partial S_v}{\partial \tau_1}, \frac{\partial S_v}{\partial \tau_2}, \frac{\partial S_v}{\partial n}, \frac{\partial S_v}{\partial t} \ldots \right) = 0,\ v = 1\ldots 15,\ k = 1\ldots 16. \qquad (26)$$

We differentiate equations (26) in the direction $\mathbf{n}$ 14 times. Then we obtain the 240 equations of the form

$$\left[ H_k\left(S_v, \frac{\partial S_v}{\partial \tau_1}, \frac{\partial S_v}{\partial \tau_2}, \frac{\partial S_v}{\partial n}, \frac{\partial S_v}{\partial t} \ldots \right) \right]_n^{(i)} = 0,\ v = 1\ldots 15,\ k = 1\ldots 16,\ i = 0\ldots 14. \qquad (27)$$

We denote

$$S_v^j = \partial^j S_v / \partial n^j, \qquad (28)$$

where $j = 0...15$, $S_v^0 = S_v$. Then 240 equations (27) can be written as:

$$O_h\left(S_v^j, \frac{\partial S_v^j}{\partial \tau_1}, \frac{\partial S_v^j}{\partial \tau_2}, \frac{\partial S_v^j}{\partial t}...\right) = 0, \; v = 1...15, \; h = 1...240, \; j = 0...15. \qquad (29)$$

We extend the variable $S_{12}^0 = S_{12} = \lambda$ along the boundary of the body as follow:

$$(\boldsymbol{\alpha} \cdot \mathbf{n}) + \lambda(\boldsymbol{\omega} + \nabla \times \boldsymbol{\alpha}) \cdot \mathbf{n} = 0. \qquad (30)$$

Let $g_0(\mathbf{r}_0, t) = 0$ be the equation of the solid surface in the coordinate system $(\mathbf{r}_0, t)$ (15). Then from (6) and (30) its speed is equal to

$$V_0 = \frac{g_{0t}}{|\nabla_0 g_0|} = \frac{g_t + ((\mathbf{u} + \boldsymbol{\alpha} + \lambda(\boldsymbol{\omega} + \nabla \times \boldsymbol{\alpha})) \cdot \nabla g)}{|\nabla_0 g_0|} =$$

$$= \frac{|\nabla g|}{|\nabla_0 g_0|}\left(V + u_n + (\boldsymbol{\alpha} \cdot \mathbf{n}) + \lambda(\boldsymbol{\omega} + \nabla \times \boldsymbol{\alpha}) \cdot \mathbf{n}\right) = 0. \qquad (31)$$

In the coordinate system $(\mathbf{r}_0, t)$ the solid body is not moving.

Thus, the dependence of the derivative to $\mathbf{n}$ in the transformed system of differential equations (29) is missing, and we obtain a closed system of 240 surface differential equations strictly along the boundary of a solid and a corresponding number of variables. In fact, some of the variables are determined from (6) and (30). At the initial time, it is required to know nontrivial vorticity $\boldsymbol{\omega}_0(\mathbf{r}_0)$ and some of its spatial derivatives on the surface (formula (20), (21)). It is always possible to find where the vorticity is non-trivial, because there is a viscous boundary layer on the surface.

The advantage is that the system actually describes the process of flow past the body surface in terms of the surface itself. Calculations show [15, 16] that the expressions (29) obtained in an analytical form are extremely cumbersome. Their writing out is beyond the scope of this paper. However, one provides a detailed algorithm of their obtaining.

4. **Overdetermination of the complete system of hydrodynamic equations**

Complete system of equations describing the hydrodynamic environment in a non-uniform unsteady 3D case is [1]

$$\frac{\partial \rho}{\partial t} + \text{div}(\rho \mathbf{u}) = 0, \qquad (32)$$

$$\frac{\partial \mathbf{u}}{\partial t} - \mathbf{u} \times \boldsymbol{\omega} = -\nabla\left(\frac{\mathbf{u}^2}{2}\right) - \frac{\nabla P}{\rho} + \frac{\nabla \sigma}{\rho} + \frac{\mathbf{F}}{\rho}, \tag{33}$$

$$\boldsymbol{\omega} = \nabla \times \mathbf{u}, \tag{34}$$

$$\psi = \text{div}\,\mathbf{u}, \tag{35}$$

$$\rho T\left[\frac{\partial s}{\partial t} + (\mathbf{u}\nabla)s\right] = Q - \text{div}\,\mathbf{q} + \Phi, \tag{36}$$

$$T = T(\rho, s),\ P = P(\rho, s), \tag{37}$$

$$\nabla w = T \cdot \nabla s + \frac{\nabla P}{\rho}, \tag{38}$$

where $\rho$, $\mathbf{u}$, $P$, $T$, $w$ и $s$ - are density, velocity, pressure, temperature, specific heat function, and the entropy (per unit mass) of the medium, respectively. Here $\mathbf{F}$ и $Q$ - are volumetric force and heat generation respectively, $\mathbf{q} = -\kappa \nabla T$ - is the heat flow equation ($\kappa$ - thermal conductivity), $\sigma$ - the viscous stress tensor, for which we have

$$\nabla \sigma = \mu\left(\Delta \mathbf{u} + \frac{1}{3}\nabla(\nabla \mathbf{u})\right) = \mu\left(-\nabla \times \boldsymbol{\omega} + \frac{4}{3}\nabla \psi\right). \tag{39}$$

Here is $\mu$ - shear dynamic viscosity. The dissipation function $\Phi$ in (36) is equal to

$$\Phi = \frac{\mu}{2}\left(\frac{\partial u_i}{\partial x_k} + \frac{\partial u_k}{\partial x_i}\right)^2 - \frac{2}{3}\mu\left(\frac{\partial u_l}{\partial x_l}\right)^2. \tag{40}$$

Equations (37) and (38) –are thermodynamic relations that depend on the properties of the medium.

We introduce a "correction" $\boldsymbol{\alpha}(\mathbf{r},t)$ to the vector $\mathbf{u}(\mathbf{r},t)$ and "pseudo-density" $\rho^\bullet(\mathbf{r},t)$, which we determine from equations:

$$\frac{\partial \rho^\bullet}{\partial t} + (\mathbf{u}+\boldsymbol{\alpha})\cdot\nabla\rho^\bullet + \rho^\bullet \text{div}(\mathbf{u}+\boldsymbol{\alpha}) = 0, \tag{41}$$

$$\frac{\partial \boldsymbol{\alpha}}{\partial t} = [\boldsymbol{\alpha}\times\boldsymbol{\omega} + \mathbf{u}\times(\nabla\times\boldsymbol{\alpha}) + \boldsymbol{\alpha}\times(\nabla\times\boldsymbol{\alpha})] - T\nabla s - \frac{\nabla\sigma}{\rho} - \frac{\mathbf{F}}{\rho}. \tag{42}$$

Termwise folded formula (42) and (33) and taking the operation «$\nabla\times$» on both sides, we find, using (38), that equation (33) is equivalent to the following expressions

$$\frac{\partial(\mathbf{u}+\boldsymbol{\alpha})_x}{\partial t} = [(\mathbf{u}+\boldsymbol{\alpha})\times(\boldsymbol{\omega}+\nabla\times\boldsymbol{\alpha})]_x - \frac{\partial}{\partial x}\left(\frac{\mathbf{u}^2}{2}\right) - \frac{\partial}{\partial x}w, \tag{43}$$

$$\frac{\partial(\boldsymbol{\omega}+\nabla\times\boldsymbol{\alpha})}{\partial t}=\nabla\times[(\mathbf{u}+\boldsymbol{\alpha})\times(\boldsymbol{\omega}+\nabla\times\boldsymbol{\alpha})]. \tag{44}$$

We go one to one to the pseudo Lagrangian variables of initial position of the gas particles and time $(\mathbf{r}_0, t)$ (i.e. marking)

$$\frac{d\mathbf{r}}{dt}=\mathbf{u}(\mathbf{r},t)+\boldsymbol{\alpha}(\mathbf{r},t)+\lambda(\boldsymbol{\omega}(\mathbf{r},t)+\nabla\times\boldsymbol{\alpha}(\mathbf{r},t)),$$

$$\mathbf{r}=\mathbf{r}(\mathbf{r}_0,t) \text{ и } \mathbf{r}_0=\mathbf{r}_0(\mathbf{r},t), \ \mathbf{r}_0|_{t=0}=\mathbf{r}. \tag{45}$$

We chose $\lambda(\mathbf{r},t)$ so that

$$\lambda(\boldsymbol{\omega}+\nabla\times\boldsymbol{\alpha})\cdot\nabla\rho^{\bullet}+\rho^{\bullet}(\boldsymbol{\omega}+\nabla\times\boldsymbol{\alpha})\cdot\nabla\lambda=0. \tag{46}$$

Hence, it can be shown that

$$\frac{\partial\mathbf{r}_0}{\partial t}+(\mathbf{u}+\boldsymbol{\alpha}+\lambda(\boldsymbol{\omega}+\nabla\times\boldsymbol{\alpha}))\cdot\nabla\mathbf{r}_0=0. \tag{47}$$

Equations (41) and (44) can be transformed to the form (see Appendix A)

$$\frac{(\boldsymbol{\omega}+\nabla\times\boldsymbol{\alpha})\cdot\nabla\mathbf{r}_0}{\rho^{\bullet}}=\frac{(\boldsymbol{\omega}_0(\mathbf{r}_0)+\nabla\times\boldsymbol{\alpha}_0(\mathbf{r}_0))}{\rho^{\bullet}_0(\mathbf{r}_0)}, \tag{48}$$

$$\frac{1}{\rho^{\bullet}}\frac{\partial(x_0,y_0,z_0)}{\partial(x,y,z)}=\frac{1}{\rho^{\bullet}_0(\mathbf{r}_0)}, \tag{49}$$

where $\boldsymbol{\alpha}_0$, $\rho^{\bullet}_0$, $\boldsymbol{\omega}_0=\nabla\times\mathbf{u}_0$ - initial distribution of the variables $\boldsymbol{\alpha}$, $\rho^{\bullet}$, $\boldsymbol{\omega}$. Let $\boldsymbol{\alpha}_0\equiv 0$ and $\rho^{\bullet}_0\equiv\rho_0$. Then it can be shown [14], equations (48) - (49) can be transformed to

$$\omega_x+\nabla\times\alpha_x-\left|\frac{\partial\mathbf{r}_0}{\partial y}\ \frac{\partial\mathbf{r}_0}{\partial z}\ \boldsymbol{\omega}_0(\mathbf{r}_0)\right|=0, \tag{50}$$

$$\omega_y+\nabla\times\alpha_y-\left|\frac{\partial\mathbf{r}_0}{\partial z}\ \frac{\partial\mathbf{r}_0}{\partial x}\ \boldsymbol{\omega}_0(\mathbf{r}_0)\right|=0, \tag{51}$$

$$\omega_z+\nabla\times\alpha_z-\left|\frac{\partial\mathbf{r}_0}{\partial x}\ \frac{\partial\mathbf{r}_0}{\partial y}\ \boldsymbol{\omega}_0(\mathbf{r}_0)\right|=0. \tag{52}$$

and the formula (49) takes the form

$$\frac{1}{\rho^{\bullet}}\frac{\partial(x_0,y_0,z_0)}{\partial(x,y,z)}=\frac{1}{\rho_0(\mathbf{r}_0)}. \tag{53}$$

As a result, we get the following overdetermined system of 21 equations (32) - (38) (41) - (43), (46), (50), (51), (53), plus any of the equations from (47) and of 20 unknowns $\rho$, $\rho^{*}$, $\mathbf{u}$, $\boldsymbol{\omega}$, $\boldsymbol{\alpha}$, $P$, $T$, $s$, $w$, $\psi$, $\lambda$, $\mathbf{r}_0$.

## 5. Three-dimensional general stream

We derive a system of equations describing the unsteady flow around the fixed rigid body in three-dimensional stream in the general case. Let

$S_1 = \alpha_x$, $S_2 = \alpha_y$, $S_3 = \alpha_z$, $S_4 = u_x$, $S_5 = u_y$, $S_6 = u_z$, $S_7 = \omega_x$, $S_8 = \omega_y$, $S_9 = \omega_z$, $S_{10} = P$, $S_{11} = \rho$, $S_{12} = \lambda$, $S_{13} = x_0$, $S_{14} = y_0$, $S_{15} = z_0$, $S_{16} = \rho^\bullet$, $S_{17} = T$, $S_{18} = s$, $S_{19} = \psi$, $S_{20} = w$, $S_{21} = q_x$, $S_{22} = q_y$, $S_{23} = q_z$. (54)

Then the system of 24 partial differential equations of the first order (32) - (38) (41) - (43), (46), (50), (51), (53), plus any of the equations from (47) can be written as

$$H_k\left(S_v, \frac{\partial S_v}{\partial \mathbf{r}}, \frac{\partial S_v}{\partial t}...\right) = 0, \quad v = 1...23, \quad k = 1...24. \quad (55)$$

We now turn to the coordinate system $(\boldsymbol{\tau}_1, \boldsymbol{\tau}_2, \mathbf{n})$ at point $M$ (see Fig. 1). Then equations (55) can be written as

$$H_k\left(S_v, \frac{\partial S_v}{\partial \tau_1}, \frac{\partial S_v}{\partial \tau_2}, \frac{\partial S_v}{\partial n}, \frac{\partial S_v}{\partial t}...\right) = 0, \quad v = 1...23, \quad k = 1...24. \quad (56)$$

Differentiating equations (56) in the direction $\mathbf{n}$ 22 times, we obtain 552 equations of the form

$$\left[H_k\left(S_v, \frac{\partial S_v}{\partial \tau_1}, \frac{\partial S_v}{\partial \tau_2}, \frac{\partial S_v}{\partial n}, \frac{\partial S_v}{\partial t}...\right)\right]_n^{(i)} = 0, \quad v = 1...23, \quad k = 1...24, \quad i = 0...22. \quad (57)$$

Denote

$$S_v^j = \partial^j S_v / \partial n^j, \quad (58)$$

where $j = 0...23$, $S_v^0 = S_v$. Then 552 equations (57) can be written as:

$$O_h\left(S_v^j, \frac{\partial S_v^j}{\partial \tau_1}, \frac{\partial S_v^j}{\partial \tau_2}, \frac{\partial S_v^j}{\partial t}...\right) = 0, \quad v = 1...23, \quad h = 1...552, \quad j = 0...23. \quad (59)$$

We extend the variable $S_{12}^0 = S_{12} = \lambda$ along the boundary of the body as follows:

$$(\boldsymbol{\alpha} \cdot \mathbf{n}) + \lambda(\boldsymbol{\omega} + \nabla \times \boldsymbol{\alpha}) \cdot \mathbf{n} = 0. \quad (60)$$

Let $g_0(\mathbf{r}_0, t) = 0$ be the equation of the solid surface in the coordinate system $(\mathbf{r}_0, t)$ (45). Similarly as in section 3, in the coordinate system $(\mathbf{r}_0, t)$, the solid is static.

Similarly, dependence on derivative to $\mathbf{n}$ in the transformed system of differential equations (59) is missing, and we obtain a closed system of 552 surface differential equations strictly along the boundary of a solid and the same number of variables (58). In fact, some variables were already defined by (6) and (60), as well as by conditions of the heat transfer on surface of the body.

## Conclusion

In this paper, using a special transformation of variables we have reduced the full set of hydrodynamic equations describing the unsteady flow around a rigid body in terms of three-dimensional flow to a system of equations on the surface. These equations can greatly simplify the numerical simulation and explore the deeper features of the flow process. First, they reduce the dimension of the task by a unit; there is no need to solve the hydrodynamic equations in the boundary layer. Second, besides the gas velocity at the surface they allow defining how all other parameters characterizing the flow (such as, $\mathbf{u}$, $P$, $\boldsymbol{\omega}$ etc.) change at the boundary.

It should be noted that not all of the obtained equations are with respect to time distinct systems. Therefore, in order to determine the stress distribution in addition to the initial data one should require examining boundary conditions. Through the previous, the information on the external flow is undoubtedly affecting the evolution of the whole effect of the flow on the body. They may need to be defined numerically using the additional code in space, the accuracy of which is important, but not in the entire volume, and only near the curves on the surface of the body along which the boundary conditions are determined.

The systems of equations obtained in this paper are essentially three-dimensional ones, ie, degenerate into the two-dimensional case. In the two-dimensional case (see Fig. 2) in order to override the hydrodynamic equations, it is necessary instead of transformations (15), (45) to use a different transformation similar to (A.12) (see Appendix A):

$$\frac{d\mathbf{r}}{dt} = \mathbf{u} + \boldsymbol{\alpha} + \frac{(\boldsymbol{\omega} + \nabla \times \boldsymbol{\alpha})}{|\boldsymbol{\omega} + \nabla \times \boldsymbol{\alpha}|^2} \times \nabla \beta, \quad \mathbf{r} = \mathbf{r}(\mathbf{r}_0, t), \quad \mathbf{r}_0 = \mathbf{r}_0(\mathbf{r}, t), \quad \mathbf{r}_0\big|_{t=0} = \mathbf{r}. \tag{84}$$

An additional condition should be implemented instead of (16)

$$\left( \frac{(\boldsymbol{\omega} + \nabla \times \boldsymbol{\alpha})}{|\boldsymbol{\omega} + \nabla \times \boldsymbol{\alpha}|^2} \times \nabla \beta \right) \nabla \rho + \rho \nabla \beta \cdot \left( \nabla \times \frac{(\boldsymbol{\omega} + \nabla \times \boldsymbol{\alpha})}{|\boldsymbol{\omega} + \nabla \times \boldsymbol{\alpha}|^2} \right) = 0. \tag{85}$$

Parameter $\beta$ is determined so that the boundary of the body did not move in the coordinates $\mathbf{r}_0$ (see (30) and (60)).

Despite the complexity, our description can successfully be used in applications, such as less expensive calculation of lift or drag coefficient in the external wing of impingement (including supersonic) flow, as well as the walls of the heat exchangers of nuclear reactors. The Approach developed here is quite general and can be applied to other problems [14, 17]. If a more simple way is subsequently proposed to override the Navier-Stokes equations, the number of computations significantly reduced. This article may be of interest to researchers engaged in finding the analytical solutions of the equations of hydrodynamics. For example, according to the proposed method the hydrodynamic equations in the volume can be reduced to a system of equations on the surface and even get an overdetermined system of surface equations. Consequently, it can already be reduced to an overdetermined system of equations along a curve on the surface, etc. up to the analytical solution.

It should be noted that the override method itself is quite simple and already described by the authors [14] where a method of overriding the hydrodynamic equations described in Sec. 2 and Sec. 4 is provided. However, in this paper we apply it to a specific practical problem.

## Appendix A

*Integrals of motion*

Consider the Euler equations of an ideal incompressible fluid in the 3D case, in the form [1]

$$\frac{\partial \boldsymbol{\omega}}{\partial t} = \nabla \times [\mathbf{u} \times \boldsymbol{\omega}], \tag{A.1}$$

$$\text{div}\mathbf{u} = 0. \tag{A.2}$$

We also consider the Lagrangian variables (i.e. markup) $\mathbf{r}_0 = \mathbf{r}_0(\mathbf{r},t)$. Multiply both sides of (A.1) to $\nabla x_0$. Then

$$\nabla x_0 \cdot \frac{\partial \boldsymbol{\omega}}{\partial t} = \nabla x_0 \cdot \nabla \times [\mathbf{u} \times \boldsymbol{\omega}] = -\text{div}(\nabla x_0 \times [\mathbf{u} \times \boldsymbol{\omega}]) = -\text{div}(\mathbf{u}(\boldsymbol{\omega} \cdot \nabla x_0)) + \text{div}(\boldsymbol{\omega}(\mathbf{u} \cdot \nabla x_0)) =$$

$$= -\mathbf{u} \cdot \nabla (\boldsymbol{\omega} \cdot \nabla x_0) + \boldsymbol{\omega} \cdot \nabla (\mathbf{u} \cdot \nabla x_0) = -\mathbf{u} \cdot \nabla (\boldsymbol{\omega} \cdot \nabla x_0) - \boldsymbol{\omega} \cdot \nabla \left(\frac{\partial x_0}{\partial t}\right).$$

Here we have used the property of Lagrange variables:

$$\frac{\partial x_0}{\partial t} + u_x \frac{\partial x_0}{\partial x} + u_y \frac{\partial x_0}{\partial y} + u_z \frac{\partial x_0}{\partial z} = 0.$$

Consequently,

$$\frac{\partial}{\partial t}(\boldsymbol{\omega} \cdot \nabla x_0) + \mathbf{u} \cdot \nabla (\boldsymbol{\omega} \cdot \nabla x_0) = \frac{d}{dt}(\boldsymbol{\omega} \cdot \nabla x_0) = 0 \Rightarrow \boldsymbol{\omega} \cdot \nabla x_0 = \omega_{0x}(x_0, y_0, z_0),$$

where $d/dt = \partial/\partial t + (\mathbf{u} \cdot \nabla)$ - time derivative in the Lagrangian variables, $\boldsymbol{\omega}_0(x_0, y_0, z_0)$ - initial vorticity distribution $\boldsymbol{\omega}$. Similarly $\boldsymbol{\omega} \cdot \nabla y_0 = \omega_{0y}(x_0, y_0, z_0)$ and $\boldsymbol{\omega} \cdot \nabla z_0 = \omega_{0z}(x_0, y_0, z_0)$.

From the continuity equation (A.2) one follows

$$\text{div}\mathbf{u} = 0 \Rightarrow \left( \frac{\partial(u_x, y, z)}{\partial(x, y, z)} + \frac{\partial(x, u_y, z)}{\partial(x, y, z)} + \frac{\partial(x, y, u_z)}{\partial(x, y, z)} \right) = 0,$$

$$\frac{1}{\Delta} \cdot \frac{d\Delta}{dt} = 0,$$

т.е. $\dfrac{\partial(x, y, z)}{\partial(x_0, y_0, z_0)} = 1$. \hfill (A.3)

Expressing $\boldsymbol{\omega}$ using (A.3), we also have $\omega_x(x_0, y_0, z_0, t) = \boldsymbol{\omega}_0 \cdot \nabla_0 x$, $\omega_y(x_0, y_0, z_0, t) = \boldsymbol{\omega}_0 \nabla_0 y$, $\omega_z(x_0, y_0, z_0, t) = \boldsymbol{\omega}_0 \cdot \nabla_0 z$, где $\nabla_0 = (\partial/\partial x_0, \partial/\partial y_0, \partial/\partial z_0)$ (see [1] page 31).

We also consider the general equation of continuity

$$\frac{\partial \rho}{\partial t} + \mathbf{u} \cdot \nabla \rho + \rho \cdot \text{div}\mathbf{u} = 0. \tag{A.4}$$

Passing to the Lagrangian variables, we find [6]

$$\frac{d\rho}{dt} + \rho \cdot \left( \frac{\partial(u_x, y, z)}{\partial(x, y, z)} + \frac{\partial(x, u_y, z)}{\partial(x, y, z)} + \frac{\partial(x, y, u_z)}{\partial(x, y, z)} \right) = 0,$$

$$\frac{d\rho}{dt} + \frac{\rho}{\Delta} \cdot \frac{d\Delta}{dt} = 0,$$

i.e.

$$\rho \cdot \Delta = \rho \cdot \frac{\partial(x, y, z)}{\partial(x_0, y_0, z_0)} = \rho_0(x_0, y_0, z_0) \text{ - integral of motion.}$$

In the case $\rho \neq const$, but $s = const$, there is a clear connection between the pressure and density $P = P_S(\rho)$ the corresponding integrals are

$$\frac{\boldsymbol{\omega} \cdot \nabla x_0}{\rho}, \quad \frac{\boldsymbol{\omega} \cdot \nabla y_0}{\rho} \text{ и } \frac{\boldsymbol{\omega} \cdot \nabla z_0}{\rho}.$$

Similarly, we obtain

$$\frac{\boldsymbol{\omega}(\mathbf{r}_0, t)}{\rho} = \frac{\boldsymbol{\omega}_0 \cdot \nabla_0 \mathbf{r}}{\rho_0},$$

The same results can be obtained if a more general transformation of the variables is used instead of Lagrangian variables

$$\frac{d\mathbf{r}}{dt} = \mathbf{u}(\mathbf{r},t) + \lambda(t)\cdot\boldsymbol{\omega}(\mathbf{r},t), \tag{A.5}$$

where $\mathbf{r} = \mathbf{r}(\mathbf{r}_0,t)$ и $\mathbf{r}_0 = \mathbf{r}_0(\mathbf{r},t)$, $\mathbf{r}_0|_{t=0} = \mathbf{r}$, а $\lambda(t)$ - some function on $t$. Similar arguments can be established that in these variables, $\boldsymbol{\omega}\cdot\nabla x_0$, $\boldsymbol{\omega}\cdot\nabla y_0$, $\boldsymbol{\omega}\cdot\nabla z_0$ and the Jacobian of this transformation $\Delta$ do not depend explicitly on time in an ideal fluid. Instead of $\lambda(t)$ we can take $\lambda(x,y,z,t)$, including $\boldsymbol{\omega}(x,y,z,t)\cdot\nabla\lambda(x,y,z,t) = 0$ to satisfy $\mathrm{div}(\mathbf{u}+\lambda\boldsymbol{\omega}) = 0$. Expressing also $\boldsymbol{\omega}$, we obtain $\omega_x(\mathbf{r}_0,t) = \boldsymbol{\omega}_0\cdot\nabla_0 x$, $\omega_y(\mathbf{r}_0,t) = \boldsymbol{\omega}_0\cdot\nabla_0 y$, $\omega_z(\mathbf{r}_0,t) = \boldsymbol{\omega}_0\cdot\nabla_0 z$.

In the case $\rho \neq const$, but $s = const$, there is a clear connection between the pressure and density $P = P_S(\rho)$ it is possible for the same reason in (A.5) to take $\lambda(x,y,z,t)$, including $\lambda\boldsymbol{\omega}\cdot\nabla\rho + \rho\boldsymbol{\omega}\cdot\nabla\lambda = 0$. Then the corresponding integrals are

$$\frac{\boldsymbol{\omega}\cdot\nabla x_0}{\rho} = \frac{\omega_{0x}(\mathbf{r}_0)}{\rho_0(\mathbf{r}_0)}, \quad \frac{\boldsymbol{\omega}\cdot\nabla y_0}{\rho} = \frac{\omega_{0y}(\mathbf{r}_0)}{\rho_0(\mathbf{r}_0)}, \quad \frac{\boldsymbol{\omega}\cdot\nabla z_0}{\rho} = \frac{\omega_{0z}(\mathbf{r}_0)}{\rho_0(\mathbf{r}_0)} \text{ и } \rho\Delta = \rho_0(\mathbf{r}_0). \tag{A.6}$$

In fact, it requires only a negligible term $\nabla\rho\times\nabla P/\rho^2$ in equation (A.1), taking into account the compressibility of the fluid, which is performed at a characteristic velocity of its motion being much less than the speed of sound. [1].

Going further. We solve the equation for the vector $\mathbf{F}$

$$\nabla\times[\mathbf{F}\times\boldsymbol{\omega}] = 0. \tag{A.7}$$

Consequently,

$$[\mathbf{F}\times\boldsymbol{\omega}] = \nabla\beta.$$

Then

$$\boldsymbol{\omega}\nabla\beta = 0 \text{ и } \boldsymbol{\omega}\times[\mathbf{F}\times\boldsymbol{\omega}] = \boldsymbol{\omega}\times\nabla\beta. \tag{A.8}$$

The solution of this equation has the form

$$\mathbf{F} = \frac{\boldsymbol{\omega}}{|\boldsymbol{\omega}|^2}\times\nabla\beta + \lambda\boldsymbol{\omega}. \tag{A.9}$$

We now change the variables

$$\frac{d\mathbf{r}}{dt} = \mathbf{u} + \frac{\boldsymbol{\omega}}{|\boldsymbol{\omega}|^2}\times\nabla\beta + \lambda\boldsymbol{\omega}, \quad \mathbf{r} = \mathbf{r}(\mathbf{r}_0,t), \quad \mathbf{r}_0 = \mathbf{r}_0(\mathbf{r},t), \quad \mathbf{r}_0|_{t=0} = \mathbf{r} \tag{A.10}$$

and thus implement the conditions

$$\boldsymbol{\omega}\nabla\beta = 0 \text{ и } \operatorname{div}\left(\frac{\boldsymbol{\omega}}{|\boldsymbol{\omega}|^2}\times\nabla\beta + \lambda\boldsymbol{\omega}\right) = \nabla\beta\cdot\left(\nabla\times\frac{\boldsymbol{\omega}}{|\boldsymbol{\omega}|^2}\right) + (\boldsymbol{\omega}\nabla)\lambda = 0. \tag{A.11}$$

Then these variables $\boldsymbol{\omega}\cdot\nabla x_0$, $\boldsymbol{\omega}\cdot\nabla y_0$, $\boldsymbol{\omega}\cdot\nabla z_0$ and the Jacobian of this transformation $\Delta$ do not depend explicitly on time. Similarly, one can take into account compressibility.

In the two-dimensional case, the change of variables (A.10) has the form

$$\frac{d\mathbf{r}}{dt} = \mathbf{u} + \frac{\boldsymbol{\omega}}{|\boldsymbol{\omega}|^2}\times\nabla\beta, \quad \mathbf{r} = \mathbf{r}(\mathbf{r}_0, t), \quad \mathbf{r}_0 = \mathbf{r}_0(\mathbf{r}, t), \quad \mathbf{r}_0\big|_{t=0} = \mathbf{r}, \tag{A.12}$$

where

$$\frac{\boldsymbol{\omega}}{|\boldsymbol{\omega}|^2}\times\nabla\beta = \frac{1}{\omega}\begin{pmatrix}-\dfrac{\partial\beta}{\partial y}\\[4pt]\dfrac{\partial\beta}{\partial x}\end{pmatrix}.$$

Condition (A.11) becomes

$$\frac{\partial\omega}{\partial x}\frac{\partial\beta}{\partial y} - \frac{\partial\omega}{\partial y}\frac{\partial\beta}{\partial x} = 0 \qquad \text{or} \qquad \beta = G(\omega, t). \tag{A.13}$$

As a result, the change of variables (A.12) can be written as

$$\frac{d\mathbf{r}}{dt} = \mathbf{u} + \mu(\omega, t)\nabla\times\boldsymbol{\omega}, \quad \mathbf{r} = \mathbf{r}(\mathbf{r}_0, t), \quad \mathbf{r}_0 = \mathbf{r}_0(\mathbf{r}, t), \quad \mathbf{r}_0\big|_{t=0} = \mathbf{r}.$$

(A.14)

These variables allow

$$\omega = \omega_0(x_0, y_0) \text{ and } \frac{\partial(x_0, y_0)}{\partial(x, y)} = 1.$$

**FIGURES**

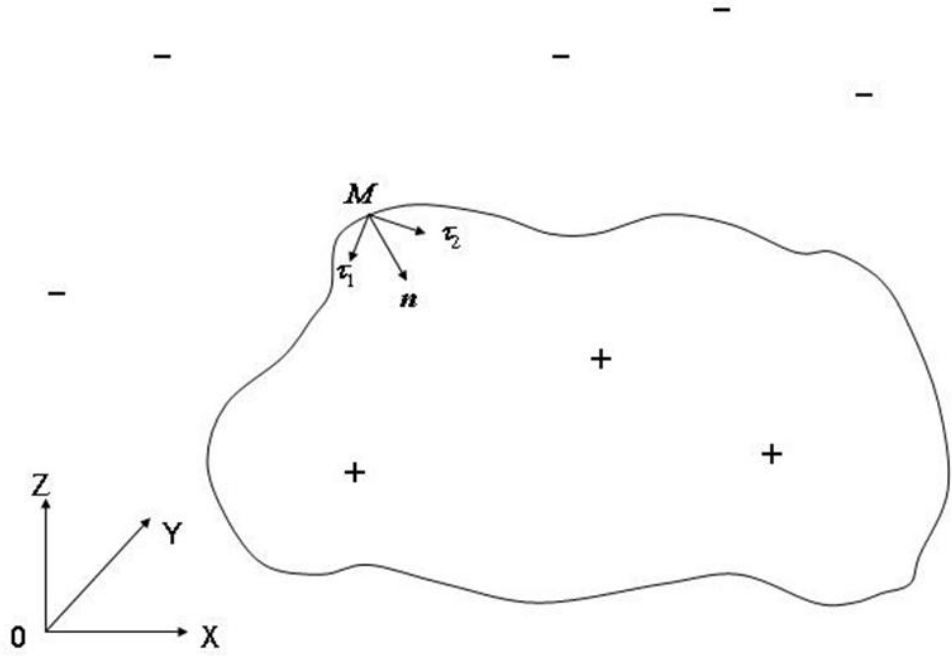

Fig. 1. Rigid body in a viscous medium

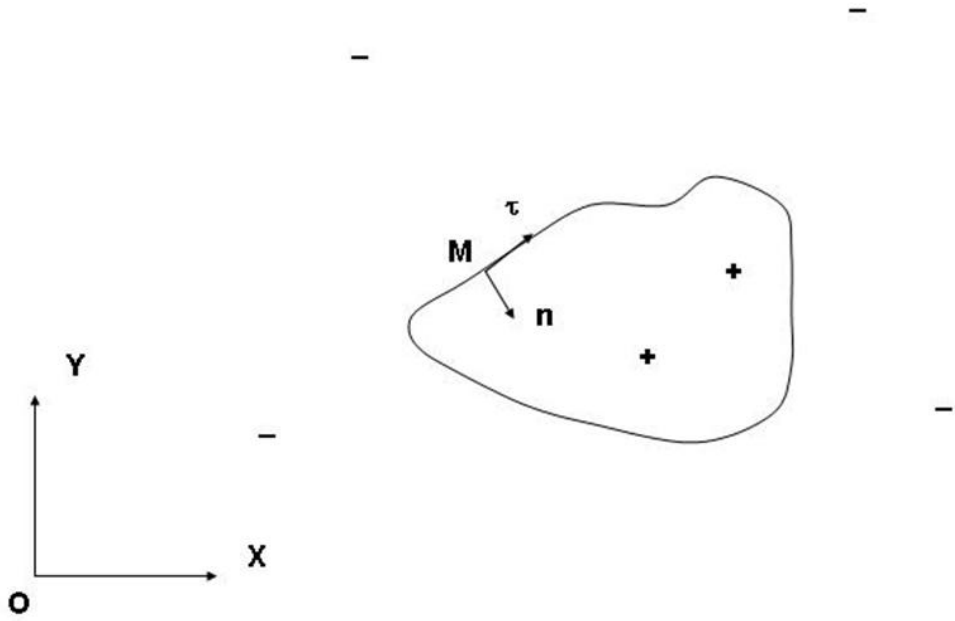

Fig. 2. A solid in viscous fluid in the two-dimensional case